\documentclass[aps,pre,reprint,amsmath,amssymb,groupaddress,superscriptaddress]{revtex4-2}

\usepackage[utf8]{inputenc}
\usepackage{amsmath}
\usepackage{amsfonts}
\usepackage{amssymb}
\usepackage{bm}
\usepackage{color}
\usepackage{graphicx}
\usepackage{xcolor}
\usepackage{mathtools}
\usepackage{physics}

\begin{document}
\title{Universal thermokinetic decomposition of short-time information fluctuations
}
\author{Giorgio Nicoletti}
\affiliation{Quantitative Life Sciences section, The Abdus Salam International Center for Theoretical Physics (ICTP), Trieste, Italy}
\affiliation{Theoretical Sciences Visiting Program, Okinawa Institute of Science and Technology Graduate University, Onna, 904-0495, Japan}
\author{Daniel Maria Busiello}
\affiliation{Theoretical Sciences Visiting Program, Okinawa Institute of Science and Technology Graduate University, Onna, 904-0495, Japan}
\affiliation{Department of Physics and Astronomy, University of Padova, Italy}
\affiliation{Max Planck Institute for the Physics of Complex Systems, Dresden, Germany}

\begin{abstract}
    \noindent Biological, artificial, and physical systems dissipate energy to accurately transmit information. While tools of information theory have been used to characterize information-processing capabilities, how reliably this information is acquired along individual trajectories, and which aspects require a thermodynamic cost, is an open question. In this work, we focus on the stochastic predictability of an arbitrary Langevin dynamics, defined as the pointwise mutual information between the current and future states of a system. We show that the fluctuations of predictability obey a universal thermokinetic decomposition at short times, which reveals that information fluctuations are suppressed by energy dissipation and become stronger with increased dynamical activity. Remarkably, we find that the average predictability, i.e., the short-time mutual information, does not carry any dependence on the underlying thermodynamic and kinetic features. Thus, the role of dissipation at short times is not to enhance information, but to reduce its fluctuations. Such dissipative control is effective only when instantiated by nonlinear operations. Moreover, energy consumption governs short- and long-time precision in stochastic oscillators through structurally different mechanisms that can be independently tuned. Our decomposition offers a fundamental thermodynamic basis for understanding the reliability of information transmission in nonequilibrium systems, the constraints on precision in biological systems, and the design of energy-limited control strategies.
\end{abstract}

\maketitle

\noindent Accurate responses typically require energy consumption \cite{bialek2005physical, lan2012energy, govern2014energy, mora2019physical, horowitz2020thermodynamic}. Energetic constraints shape the precision of information processing across a wide range of biological contexts, from sensing architectures \cite{nicoletti2025optimal, sartori2014thermodynamic, tabanera2025multiple} to chemical oscillators \cite{cao2015free, barato2017coherence, santolin2025dissipation, zhang2020energy}. The natural framework to quantify the accuracy of information processing is provided by information-theoretic quantities, from mutual information between input and output signals \cite{cheong2011information, tkavcik2016information, bauer2023information, nicoletti2024tuning,tkavcik2025information} to transfer entropy \cite{schreiber2000measuring, horowitz2014second, das2025exact} and information rate \cite{tostevin2009mutual, reinhardt2023path, moor2023dynamic, ripsman2025using, leighton2025flow}. Recent work in particular has emphasized the importance of short-time response, focusing on how the internal degrees of freedom of chemotactic agents encode external signals over short timescales \cite{mattingly2021escherichia, rode2024information, nakamura2026quantification}. Yet, drawing general connections between information and dissipation is a long-standing problem, leaving open the question of whether thermodynamic activity leaves a quantifiable and universal signature in information processing.



Here, we address this question for any system obeying a Langevin dynamics by studying the pointwise mutual information between successive states of a stochastic system at short times. This measure quantifies the predictability of the system's evolution, and its fluctuations along individual trajectories determine how reliably it can process information. 
We show that its average, the time-lagged mutual information, carries no thermodynamic signature -- rather, it is identical for equilibrium and nonequilibrium systems sharing the same stationary distribution. However, we find a markedly different picture for the fluctuations of information. Dissipation can actively suppress these fluctuations, driving the system toward more reliable information acquisition by overcoming the randomness introduced by noise. Dynamical activity, in contrast, can amplify them. The interplay between energy consumption and kinetic effects determines how consistently information is shared between the past and the immediate future along individual trajectories. We derive a universal decomposition of information fluctuations that precisely quantifies the competition between these thermokinetic terms, which holds independently of the details of the dynamics.

This result reveals a profound and general connection between information and dissipation that emerges only at the level of fluctuations. We apply our decomposition to different systems, highlighting the interplay between dissipation, nonlinear forces, and the system's geometry in shaping the reliability of information processing at short times. In doing so, we establish a universal thermodynamic basis that is fundamentally distinct from the dissipative mechanisms governing long-time precision.

\section{Information fluctuations along a stochastic trajectory}
\noindent We consider a stochastic system with $N$ degrees of freedom (DOFs) described by a state vector $\bm{x}_t$ at time $t$. The system evolves according to a multidimensional Langevin equation
\begin{equation}
\label{eqn:Langevin}
    \dot{\bm{x}}_t = \bm{F}(\bm{x}_t) + \sqrt{2} \,\bm{\sigma} \, \bm{\xi}_t
\end{equation}
where $\bm{F}$ is an arbitrary force field acting on the DOFs, $\bm{\xi}_t$ a vector of Gaussian white noises, and $\sigma$ determines the diffusivity $N\times N$ matrix as $\bm{D} = \bm{\sigma}^T\bm{\sigma}$.

Along a stochastic trajectory, the state $\bm{x}_t$ continuously fluctuates while evolving in time. These fluctuations reflect two intertwined properties -- the intrinsic stochasticity of the dynamics and the statistical information that each infinitesimal time step retains about the preceding state. Quantitatively, these time-dependent fluctuations can be characterized through the time-lagged pointwise mutual information (pMI):
\begin{equation}
\label{eqn:mpi}
    i_\tau = \log \frac{p(\bm{x}_{t + \tau}| \bm{x}_t)}{p(\bm{x}_{t + \tau})} \;,
\end{equation}
where $p(\bm{x}_{t + \tau}| \bm{x}_t)$ is the propagator solving the Fokker-Planck (FP) equation associated with Eq.~\eqref{eqn:Langevin}, and $p(\bm{x}_t)$ is the state probability distribution at time $t$. The pMI $i_\tau$ is a measure of statistical dependencies between the state at time $t$ and the one at time $t + \tau$, and can be interpreted as a stochastic predictability -- more simply predictability throughout this manuscript. The pMI is zero if and only if the future state is independent of the current one. However, being a fluctuating quantity, $i_\tau$ is not guaranteed to be positive. When $i_\tau \gg 0$, we have $p(\bm{x}_{t+\tau} | \bm{x}_t) \gg p(\bm{x}_{t+\tau})$, meaning the observed future state is much more likely given the current state than it is on average. When $i_\tau \ll 0$, the opposite holds -- the probability of reaching the future state after a time $\tau$ is much smaller than that of finding it by averaging over all possible current states.



The ensemble average of $i_\tau$ is the standard mutual information $I_\tau = \ev{i_\tau}$, which quantifies the average predictability of the system's state after a lag $\tau$. Therefore, $i_\tau$ has to be non-negative on average. 
A large mutual information 
identifies pairs of states that share a large predictability in terms of the information of the future contained in the present. On the other hand, a small value signals that the future evolution of the state is close to being independent of the current one. Furthermore, the stochasticity of $i_\tau$ signals how this predictability fluctuates along a stochastic trajectory. 
We quantify these fluctuations around the average by computing the variance of $i_\tau$, which we denote with $\mathcal{F}_\tau$:
\begin{equation}
    \mathcal{F}_\tau = \langle i^2_\tau \rangle - I_\tau^2 \;.
\end{equation}
Large values of $\mathcal{F}_\tau$ correspond to strong deviations from the average predictability, indicating that $i_\tau$ varies widely along the trajectory -- the system transiently visits regions of phase space that are either unusually informative or unusually uninformative about the future. Conversely, a small value of $\mathcal{F}_\tau$ signals that predictability is approximately uniform across the state space, and the average mutual information $I_\tau$ is therefore a faithful characterization of the system's behavior at each point along its trajectory. We sketch these ideas in Figure \ref{fig:sketch_OU}a.

\section{Kinetic and thermodynamic components of information fluctuations}

\subsection{Short-time moment generating function}
\noindent While the pMI and its fluctuations can be computed at any time delay $\tau$, we here focus on the short-time behavior of $i_\tau$. This limit uncovers how the statistical properties of a system's predictability emerges from the local dynamical structure, before the trajectory has time to explore the entire state space.
To derive a general characterization of the pMI, we start by computing the moment generating function (MGF) of $i_\tau$, $M_\tau(\lambda) = \ev{e^{\lambda i_\tau}}$. 

The short-time propagator of the dynamics follows a Gaussian distribution independently of the force field driving the system. Moreover, in the case of stationary processes, $p(\bm{x}_\tau) = p_\mathrm{st}(\bm{x}_\tau)$, the stationary solution of the FP equation. Therefore, up to the first order in $\tau$, the steady-state MGF takes a particularly simple form (see Methods):
\begin{align}
\label{eqn:MGF_shorttime}
    M^\mathrm{st}_\tau(\lambda) & = f^{(\lambda)} \ev{1+ \tau \frac{\lambda^2}{1-\lambda^2} \bm{\nabla} \cdot \bm{F}(\bm{x})}_{p_\mathrm{st}^{1-\lambda}} \;,
\end{align}
with $f^{(\lambda)} = (1 + \lambda)^{-N / 2}(4\pi \tau)^{-\lambda N / 2}\det(\mathbf{D})^{-\lambda/2}$. To derive this expression, we are not taking any assumption on the properties of the underlying system but stationarity. However, also this condition can be further relaxed as shown in the Supplementary Information.


\subsection{Thermokinetic decomposition}

\noindent Given the MGF, we can immediately compute the average pMI and its fluctuations as $I_\tau = \partial_\lambda M_\tau(\lambda)|_{\lambda = 0}$ and $\mathcal{F}_\tau = \partial_\lambda M_\tau(\lambda)|_{\lambda = 0}$, respectively. By exploiting the form of Eq.~\eqref{eqn:MGF_shorttime}, in the simple case of additive noise, we obtain the following form of the short-time stationary mutual information, i.e., the average predictability (see Methods):
\begin{equation}
\label{eqn:I_tau}
    I_\tau = H[p_{\rm st}] - \frac{N}{2} \left( 1 + \log(4 \pi \tau (\det\bm{D})^{1/N})\right)
\end{equation}
where $H[p_\mathrm{st}]$ is the differential entropy of the stationary distribution $p_\mathrm{st}$. This entropic term connects the information shared between current and future state of a system to the intrinsic uncertainty of the steady-state distribution. The second term in Eq.~\eqref{eqn:I_tau}, on the other hand, encodes the effect of additive noise and depends on both system dimensionality and noise correlation structure. In particular, when $\tau \to 0$, the short-time mutual information diverges logarithmically as the dynamics becomes dominated by the deterministic coupling over the noise. Importantly, $I_\tau$ depends solely on the stationary distribution and diffusion matrix, and contains no information on the underlying thermodynamic structure of the system, which is dictated by the interplay between the force field and the noise.

The situation is fundamentally different when computing predictability fluctuations, i.e., the variance of the pMI. Indeed, for a stationary process driven by additive noise and an arbitrary force field, the expression of the short-time information fluctuations reads (see Methods):
\begin{equation}
    \label{eqn:F_additive}
    \mathcal{F}_\tau = \frac{N}{2} + \mathcal{F}_H[-\log p_{\rm st}] - 2 \tau \left(\Sigma - 4\mathcal{T}\right)
\end{equation}
where
\begin{equation}
    \mathcal{F}_H[- \log p_\mathrm{st}] = \int d\bm{x} \, p_\mathrm{st}(\bm{x})\left[\log p_\mathrm{st}(\bm{x})\right]^2 - H[p_\mathrm{st}]^2
\end{equation}
is the variance of the stochastic entropy \cite{seifert2012stochastic} of $p_{\rm st}$, sometimes known as varentropy \cite{kontoyiannis2013optimal}. The first term in Eq.~\eqref{eqn:F_additive}
represents the noise contribution and depends on the system's dimensionality, while the last term encodes thermokinetic properties. Indeed, it can be written solely as a function of the entropy production rate $\Sigma$ \cite{seifert2012stochastic,peliti2021stochastic}:
\begin{equation}
    \Sigma = \int d\bm{x} ~\frac{\bm{J}_{\rm st}(\bm{x})^T \bm{D}^{-1} \bm{J}_{\rm st}(\bm{x})}{p_{\rm st}(\bm{x})} \;,
\end{equation}
where $\bm{J}_{\rm st} = -\bm{F}~p_{\rm st} + \bm{D} \cdot \bm{\nabla} p_{\rm st}$, and the traffic $\mathcal{T}$ \cite{maes2020frenesy,di2025force}:
\begin{align}
\label{eq:traffic_general}
    \mathcal{T}
    =&  \frac{1}{4} \int d\bm{x} \, p_\mathrm{st}(\bm{x}) \bm{F}(\bm{x})^T \bm{D}^{-1} \bm{F}(\bm{x}) + \nonumber \\
    & + \frac{1}{2} \int d\bm{x} \, p_\mathrm{st}(\bm{x}) \bm{\nabla}\cdot \bm{F}(\bm{x}) \; ,
\end{align}
which represents the time-symmetric part of the stochastic action. While $\Sigma$ accounts for the irreversibility of a system, $\mathcal{T}$ quantifies its dynamical activity, i.e., how often and rapidly transitions occur, independently of their directionality, and provides a kinetic proxy to estimate underlying dissipative features \cite{di2025force}. Eq.~\eqref{eqn:F_additive} constitutes the main result of this work. In the Supplementary Information, we show that it can be extended to non-stationary trajectories and multiplicative noise.

The main takeaway of Eq.~\eqref{eqn:F_additive} is that thermodynamics manifestly controls the fluctuations of predictability, while being irrelevant to determine its average value. In particular, dissipation suppresses information fluctuations by biasing the system toward directed, irreversible trajectories that exhibit reduced variability in the system's evolution at short times. Conversely, the traffic $\mathcal{T}$ encodes undirected dynamical activity that is likely to broaden the range of states that can be explored by stochastic trajectories at short times. However, kinetic activity can also counteracts the suppression induced by dissipation -- as the traffic can take negative values whenever the force field constraints the stochastic trajectories, for example through trapping mechanisms at equilibrium \cite{di2025force}. 
This decoupling between the average predictability and its fluctuations, together with the competition between dissipative and kinetic contributions, suggests that short-time precision and long-term accuracy might be controlled through very different mechanisms. More in detail, all dissipative processes that do not affect the dynamical activity decrease the fluctuations of short-time predictability, while all energy-consuming processes are equally important to drive precise operations at long times.

As shown in the Supplementary Information, $\mathcal{F}_\tau$ can be equally written in terms of the diffusion-weighted Fisher information of $p_{\rm st}$ with respect to changes in the dynamical degrees of freedom $\bm{x}$ \cite{ito2020stochastic}.


\begin{figure}
    \centering
    \includegraphics[width=\columnwidth]{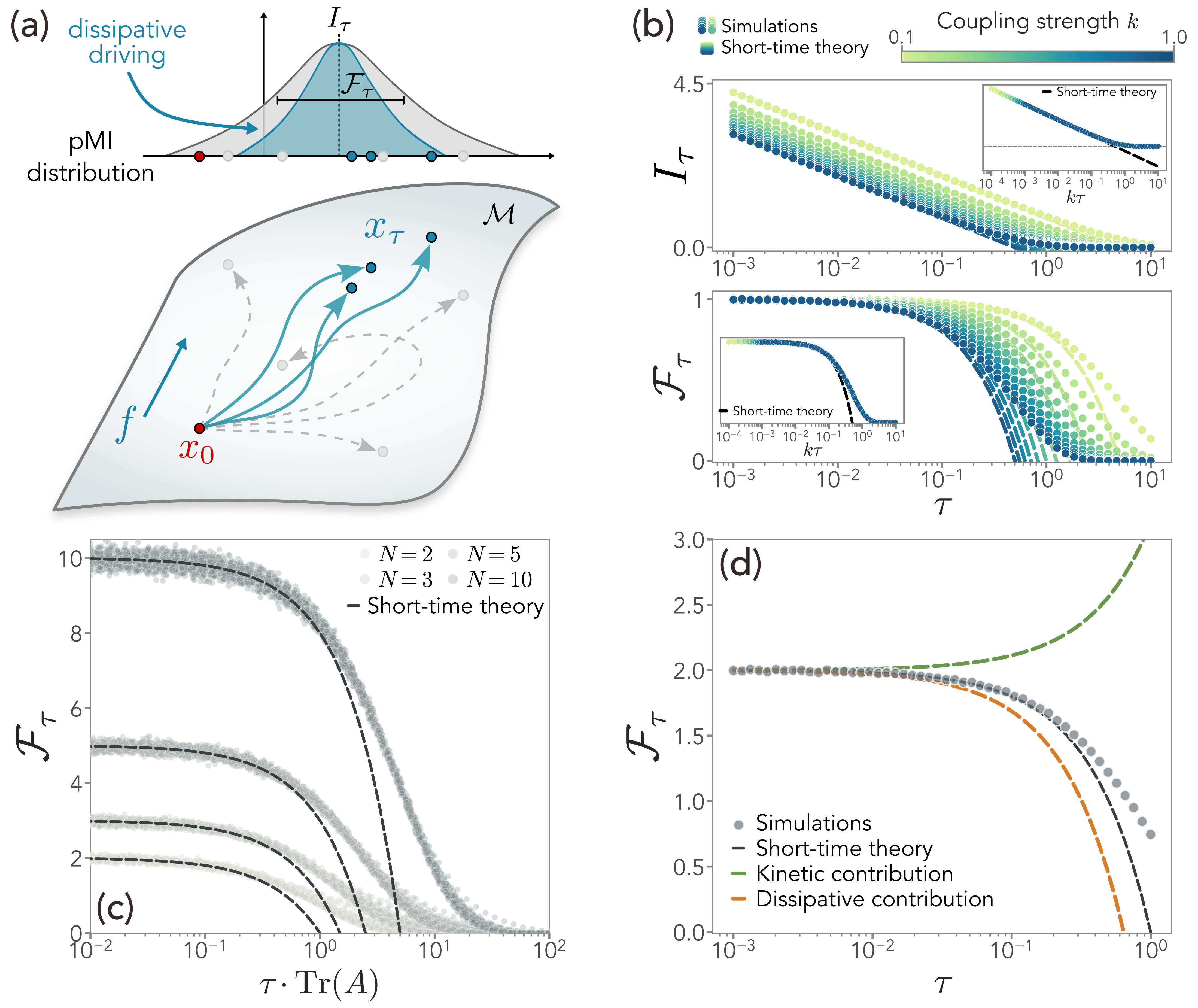}
    \caption{(a) The state of a stochastic process described by Eq.~\eqref{eqn:Langevin} evolves from $\bm{x}_0$ to $\bm{x}_\tau$, fluctuating in time and generating a set of stochastic trajectories in a manifold $\mathcal{M}$. These fluctuations shape the information shared between present and future states, which is captured by the distribution of the pointwise mutual information (pMI) between $\bm{x}_0$ and $\bm{x}_\tau$, $i_{\tau}$. While a dissipative driving $f$ does not impact the average pMI, $I_\tau$, higher dissipation is in general able to reduce the short-time fluctuations of information, $\mathcal{F}_\tau$, quantified by the variance of the pMI. (b) For a one-dimensional Ornstein-Uhlenbeck (OU) process with $F(x) = -kx$, $I_\tau$ diverges logarithmically as $\tau \to 0$, while $\mathcal{F}_\tau$ converges to $1$. Dashed lines are theoretical predictions, dots are simulations. Insets: for a $1D$ OU, both $I_\tau$ and $\mathcal{F}_\tau$ only depends on the rescaled time $k\tau$. (c) For multidimensional OU processes in $N$ dimensions, $\mathcal{F}_\tau$ saturates at $N$ for $\tau\to 0$ and its shape only depends on the rescaled time $\tau \, \mathrm{Tr}(A)$, where $\mathbf{A}$ is the drift matrix. In this panel, dots are simulations over $10^6$ realizations of random $\mathbf{A} \sim \mathcal{N}(0, (4 N)^{-1})$, and $\mathbf{D}$ is the identity. (d) Kinetic and thermodynamic contributions to $\mathcal{F}_\tau$ for a $2$-dimensional OU process out of equilibrium. The dissipative contribution tends to reduce fluctuations, while kinetic activity increases them. In this panel, $A_{11}=A_{22}=A_{21} = 1/2$, $A_{12} = -3/4$, $D_{ii}=1$ and $D_{ij} = 1/2$ for $i,j =1,2$.}
    \label{fig:sketch_OU}
\end{figure}

\subsection{Thermodynamic invariance of information fluctuations in linear systems}
\noindent We first focus on linear systems. Consider a $N$-dimensional Ornstein-Uhlenbeck (OU) process with $\bm{F}(\bm{x}_t) = \bm{A} \bm{x}_t$. In this scenario, both the deterministic drift and the noise are Gaussian, so that the varentropy reduces to a function of dimensionality alone and the inflow rate is constant (see Methods).

In one dimension, $F(x_t) = -k x_t$, where $k$ denotes the self-coupling strength, so that $k^{-1}$ sets the relaxation timescale. The average predictability in Eq.~\eqref{eqn:I_tau} simplifies to $I_\tau = -\frac{1}{2}\log(2k\tau)$, which remains positive for $\tau <(2k)^{-1}$ and is independent of the diffusion coefficient. As expected, when $\tau\to0$, the self-relaxation dominates the noise and $I_{\tau}$ diverges logarithmically. On the other hand, the short-time information fluctuations simply becomes $\mathcal{F}_\tau = 1 - 2 k \tau$. Indeed, since the entropy production vanishes for one-dimensional Gaussian systems, the predictability fluctuations are completely determined by the traffic $\mathcal{T} = -k/4$. Overall, the behavior of the short-term statistics of the PMI is fully controlled by the rescaled dimensionless timescale $k \tau$. Indeed, in Fig.~\ref{fig:sketch_OU}b, we show a collapse of $I_{\tau}$ and $\mathcal{F}_\tau$ when such rescaling is applied. Moreover, Figure \ref{fig:sketch_OU}b also shows that our analytical results remain a good approximation even at moderately large values of $k \tau$.

These results directly extend to $N$ dimensions. In this case, $\mathcal{F}_\tau = N + 2 \tau \Tr\bm A$, showing that short-time information fluctuations is fully determined by the system dimensionality and the trace of the drift matrix that encodes the timescales at play (Figure \ref{fig:sketch_OU}c). In particular, since $\Tr\bm A$ does not depend on the antisymmetric part of the interaction matrix, $\mathcal{F}_\tau$ is insensitive to non-reciprocal interactions. This is a structural consequence of linearity: the Lyapunov equation enforces an exact cancellation between thermodynamic and kinetic contributions in Eq.~\eqref{eqn:F_additive}, as shown in Figure \ref{fig:sketch_OU}d. Stated differently, in Gaussian processes, any increase in the nonequilibrium driving must be accompanied by a corresponding increase in traffic component by the same amount. Thus, dissipative control of predictability fluctuations is an intrinsic feature that might emerge solely in nonlinear systems.



\section{Dissipative control of information fluctuations}
\noindent To study how dynamical nonlinearities reveal the role of thermodynamics in constraining information fluctuations at short time, we study a simple model of a Brownian motor \cite{reimann2002brownian, hyeon2017physical}. Consider a particle diffusing on a ring under a periodic potential with a non-equilibrium driving, $F(x_t) = f - k\sin(x_t)$ with $x_t \in [0, 2\pi]$. Here, $k$ sets the height of the periodic potential barriers and $f$ acts as an energetic source that drives a net probability current around the ring. Although the system does not admit an analytic steady-state solution, the entropy production can be written exactly as $\dot\Sigma = f J / D$, where $J = \langle f - k\sin x\rangle_\mathrm{st}$ is the steady-state current (see Supplementary Information).

The presence of a nonlinear force field implies that both entropy production and traffic depend explicitly on $p_\mathrm{st}$, which is itself controlled by the driving $f$. As $f$ increases, the driving progressively homogenizes $p_\mathrm{st}$ over the ring (Figure \ref{fig:dissipative}a), simultaneously increasing the entropy production and decreasing the varentropy (Figure \ref{fig:dissipative}b-c). Both effects act in the same direction, producing a systematic suppression of $\mathcal{F}_\tau$ with increasing dissipation (Figure \ref{fig:dissipative}d). 
This suppression has a sharp limit. In the large-$f$ regime, we obtain
\begin{equation}
    \mathcal{F}_\tau \xrightarrow{f\to\infty} \frac{1}{2} + \frac{k^2}{2\dot\Sigma D} \;,
\end{equation}
which establishes a lower bound on predictability fluctuations. As the driving promotes a stronger delocalization of the particle over the ring and the entropy production grows, $\mathcal{F}_\tau$ is driven towards $1/2$, which identifies the unavoidable fluctuations due to the Gaussian white noise. Entropy production therefore does not merely reflect nonequilibrium activity -- it actively quenches information fluctuations along stochastic trajectories, driving them toward a lower bound set by the noise alone.

\subsection{Decoupled dissipative control of coherence and information fluctuations}
\noindent We now show that, when applied to stochastic oscillators, our decomposition reveals that short- and long-time precisions are independently controlled through different dissipative mechanisms. The normal form of a noisy oscillating system near a Hopf bifurcation is captured by the Stuart-Landau equations \cite{Kuramoto1984}:
\begin{eqnarray}
    \dot{r} &=& \alpha r - \beta r^3 + \sqrt{2 \epsilon} ~\xi_r \nonumber \\
    \dot{\theta} &=& \omega + \delta r^2 + \sqrt{2 \frac{\epsilon}{r^2}} \xi_\theta \;,
\end{eqnarray}
where $\xi_r$ and $\xi_\theta$ are two uncorrelated Gaussian white noises with variance $c$, $r$ is the oscillation amplitude, and $\theta$ its phase. The parameter $\alpha$ controls the distance from the bifurcation: for $\alpha>0$, the system sustains limit-cycle oscillations around the deterministic fixed point $r^* = \sqrt{\alpha/\beta}$, where $\beta$ is the nonlinear saturation coefficient that sets the amplitude of the cycle. The angular velocity $\omega$ determines the oscillation frequency, while $\delta$ couples amplitude and phase fluctuations. The noise intensity $\epsilon$ and variance $c$ jointly set the magnitude of stochastic perturbations around the limit cycle.

\begin{figure}
    \centering
    \includegraphics[width=\linewidth]{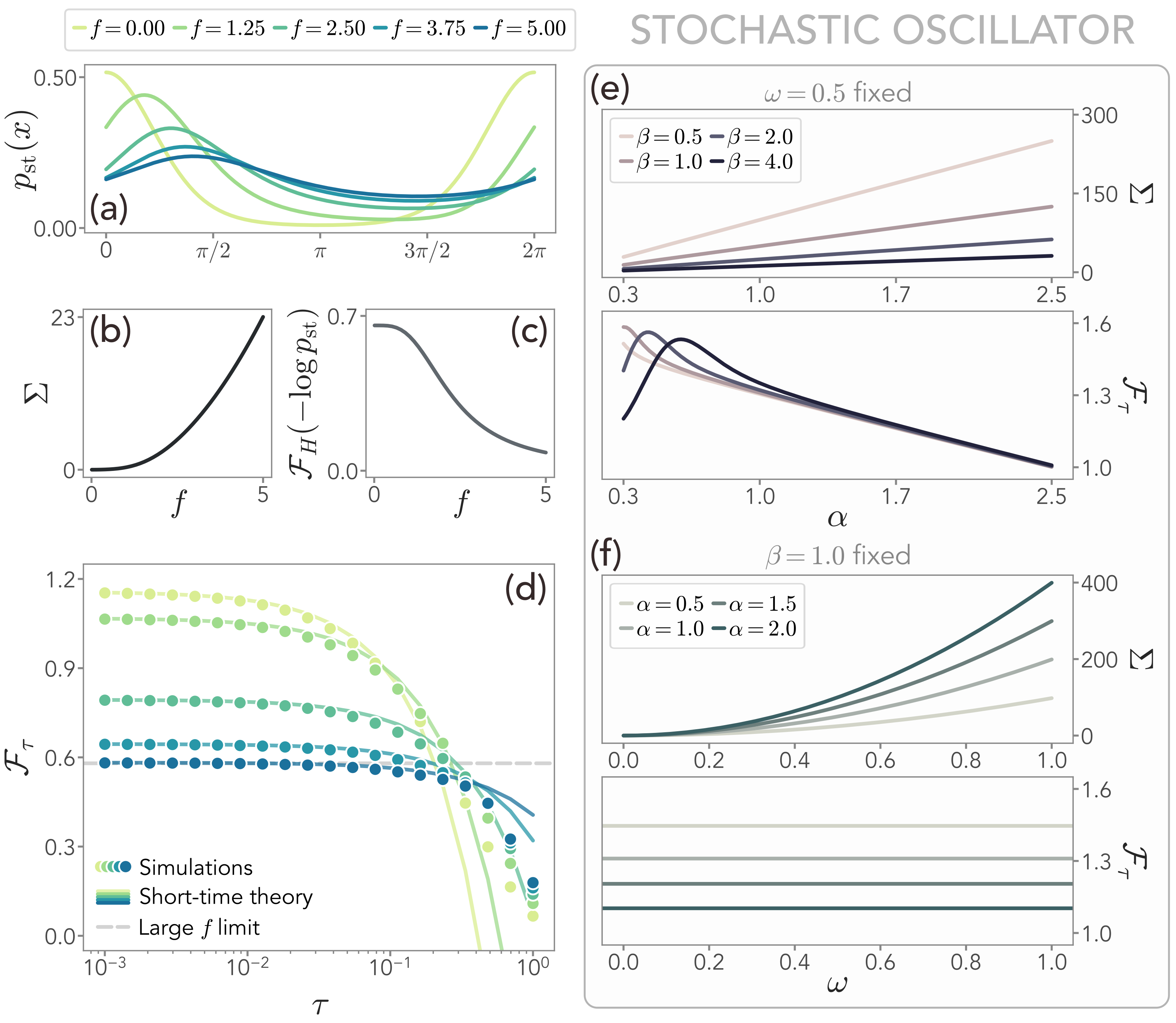}
    \caption{(a-c) Steady-state distribution $p_\mathrm{st}$ of a particle in a one-dimensional periodic potential with a nonequilibrium driving $f$. The entropy production $\Sigma$ increases with $f$, while the varentropy $\mathcal{F}_H(-\log p_\mathrm{st})$ decreases as the distribution becomes more uniform. (d) Information fluctuations $\mathcal{F}_\tau$ are suppressed by the nonequilibrium driving, up to the theoretical minimum (gray dashed line). Our short-time theory (solid lines) is well-matched by numerical simulations (circles) even for relatively large values of $\tau$. In these panels, $k = 2$ and $D = 1$. (e-f) In a stochastic oscillator, entropy production quantifies long-term precision as it is related to the number of coherent oscillations, and is controlled by both the angular velocity $\omega$ and the bifurcation parameter $\alpha$. However, information fluctuations can only be suppressed by increasing $\alpha$, showing that long-time and short-time precision are fundamentally decoupled. In these panels, $\epsilon = 0.01$ and $\delta = 0$.}
    \label{fig:dissipative}
\end{figure}

At stationarity, the probability distribution of this system is flat along $\theta$ by rotational symmetry, while the radial distribution is shaped by the competition between the deterministic confinement toward $r^*$ and the noise. Since the radial dynamics is not affected by the phase, $p_\mathrm{st}(r,\theta)$ is independent of both $\omega$ and $\delta$. Therefore, from Eq.~\eqref{eqn:F_additive}, we see that $\mathcal{F}_\tau$ inherits this independence, and is determined only by the radial dynamical parameters. In the weak-noise limit ($\epsilon \to 0$), the short-time information fluctuations take a particularly simple form:
\begin{equation}
\label{eqn:oscillator_short}
    \mathcal{F}^{\rm weak}_{\tau} = \frac{3}{2} - 2 \tau \left( 2 \alpha - 4 \epsilon c ~\frac{\beta}{\alpha}\right) \;.
\end{equation}
This quantity reflects the short-term precision of the oscillator -- increasing $\alpha$ increases dissipation along the radial component, and enhances predictability by suppressing information fluctuations (Figure \ref{fig:dissipative}e).

The picture for the long-time precision, however, is markedly different. We define the coherence $\mathcal{N}$ as the number of oscillations completed before phase correlations are lost. This quantity measures the long-time phase precision of the oscillator, and thus its ability to act as a clock. For stochastic limit cycles in the weak-noise regime, a recent study \cite{santolin2025dissipation} showed that $\mathcal{N}$ is bounded by the entropy production per cycle, $\Sigma_{\rm c}$. In this system, this bound is saturated in the isochronous case $\delta=0$, where amplitude fluctuations do not feed into phase diffusion, and in the weak-noise limit. In particular (see Supplementary Information),
\begin{equation}
\label{eqn:oscillator_long}
    \mathcal{N}^{\rm weak} = \Sigma^{\rm weak}_{\rm c} = \frac{\alpha}{\beta \epsilon c} 2 \pi \omega\;.
\end{equation}
Comparing Eq.~\eqref{eqn:oscillator_short} and Eq.~\eqref{eqn:oscillator_long} highlights that dissipation plays a fundamentally different role at short and long times. While both $\mathcal{F}_\tau$ and $\mathcal{N}$ depend on the radial parameters, the angular velocity only enters $\mathcal{N}$. Thus, increasing the entropy production through $\omega$ improves long-time coherence, but leaves the short-time information fluctuations unchanged (Figure \ref{fig:dissipative}f). Conversely, suppression of $\mathcal{F}_\tau$ through radial confinement can improve $\mathcal{N}$. This highlights a decoupling between long-time and short-time precision that is entirely controlled by rotational symmetry, revealing an intimate connection between the geometry of nonequilibrium driving and the fluctuations in system's predictability.

\section{Discussions}
\noindent In this work, we have derived a universal decomposition of predictability fluctuations, quantified by the variance of the pMI between the current and future states of a system. We uncover that such fluctuations are controlled by a contribution stemming from the geometry of the steady-state distribution, a term depending on the system's dynamical activity, and a component proportional to the entropy production rate. Our theory is valid for any stochastic process governed by Langevin equation, and can be extended to non-stationary systems (see Supplementary Information). This decomposition establishes a fundamental asymmetry -- while the average predictability does not depend on thermokinetic quantities, its fluctuations are not. At short times, dissipation does not generate information, but can rather stabilize it by suppressing the variability with which information is acquired along individual trajectories.

We have applied these ideas to exemplary systems, showing that nonlinear interactions are fundamental in allowing dissipation to affect $\mathcal{F}_\tau$. 
Furthermore, through the model of a stochastic oscillator near a Hopf bifurcation, we have shown that our decomposition reveals that dissipation may act along decoupled pathways -- one controlling predictability fluctuations, the other improving long-time coherence.
Such a decoupling is intimately tied to the geometry of nonequilibrium steady states, suggesting that the impact of energy dissipation on short-time information transmission depends on how the associated currents are organized in state space \cite{dechant2022geometric}. This observation may have direct implications for biochemical oscillators, where the reliability of individual cycles and long-time coherence are both functionally relevant, yet can in principle be tuned through distinct biological mechanisms. More generally, our results point to geometric design principles for incorporating nonequilibrium constraints so as to enhance the precision and reliability of information transmission. 

Recent work has explored analogous connections between dissipation and information at short times, but exclusively at the average level \cite{leighton2025flow, cho2025exact}. The present results uncover instead that, at short times and along single trajectories, information fluctuations are the only quantities that retain signatures of the system’s thermodynamics, highlighting their potential physical relevance beyond average quantities. This direction remains largely unexplored. Future investigations are needed to understand whether equivalent decompositions hold for jump processes and chemical master equations. Such an extension would broaden the universality class of our theory to the discrete-state models underlying most biochemical networks and would make it possible to explore how dissipation and dynamical activity shape the precision of information transmission between various observables.

A particularly promising direction is to study the underlying thermodynamic and kinetic properties of the information shared between an incoming signal and the short-time response of a coupled subset of degrees of freedom. In particular, understanding how information fluctuates when it is transferred between different variables -- and what role is played by energy consumption -- will shed more light on the link between accuracy and dissipation in biological systems, directly addressing the thermodynamic cost of reliable sensing and signal transduction. Whether the thermokinetic decomposition derived here extends beyond the present setting, and whether it allows the system to implement control strategies at limited energy budgets, are questions that will yield further universal constraints on the precision of biological information processing.

\begin{acknowledgments}
    \noindent This research was conducted while visiting the Okinawa Institute of Science and Technology (OIST) through the Theoretical Sciences Visiting Program (TSVP). The authors thank TSVP and OIST for hosting them. D.M.B. is funded by the program STARS@UNIPD with the project ``ActiveInfo''.
\end{acknowledgments}

\section{Methods}
\subsection{Short-time expansion of the propagator}
\noindent We consider the Fokker-Planck equation associated with a generic Langevin dynamics, 
\begin{equation}
    \partial_t p(\bm{x}_t) = - \bm{\nabla} \cdot \left(\bm{F}(\bm{x}_t) p(\bm{x}_t)\right) + \bm{\nabla} \cdot \bm{\nabla} \cdot \left( \bm{D}(\bm{x}_t) p(\bm{x}_t)\right)
\end{equation}
where the diffusion matrix is equal to $\bm{D}(\bm{x}_t) = \bm{\sigma}(\bm{x}_t)^T\bm{\sigma}(\bm{x}_t)$ and, in principle, may depend on $\bm{x}$. We follow the convention that the operator $\bm{\nabla} \cdot \bm{\nabla} \cdot ()$ acts on tensors as $\bm{\nabla} \cdot \bm{\nabla} \cdot\left(\bm{A}\right) = \sum_{i= 1}^N\sum_{j = 1}^N \partial_i \partial_j A_{ij}$, with $\partial_i = \partial / \partial x_i$.

When $\tau$ is much smaller than the characteristic timescale of the system evolution, the propagator from $\bm{x}_0$ to $\bm{x}_\tau$ is amenable to a short-time approximation. At the leading order,
\begin{equation}
    \bm{x}_\tau = \bm{x}_0 + \bm{F}(\bm{x}_t) \, \tau + \sqrt{2} \, \bm{\sigma} \, dW_t \;,
\end{equation}
where $dW_t$ is a Wiener process whose variance scales as $\sqrt{\tau}$. Therefore, we immediately obtain:
\begin{align}
\label{eqn:methods:short_prop}
    p(\bm{x}_\tau | \bm{x}_0) = & \frac{1}{(4 \pi \tau)^{N/2} \log\det\bm{D}(\bm{x}_t)^{1/2}} \times \\
    & \times e^{-\frac{1}{4 \tau} (\bm{x}_\tau - \bm{x}_0 - \bm{F}(\bm{x}_\tau)^T \bm{D}(\bm{x}_t)^{-1} (\bm{x}_\tau - \bm{x}_0 - \bm{F}(\bm{x}_\tau))} \;. \nonumber
\end{align}
The short-time propagator reflects the (Gaussian) statistics of the noise term. From this point on, for the sake of clarity, we use the following short-hand notation, $\bm{F}(\bm{x}_t) \to \bm{F}_x$, $\bm{D}(\bm{x}_t) \to \bm{D}_x$, $p(\bm{x}_0) \to p_0$, and $p(\bm{x}_\tau) \to p_\tau$.

\subsection{Moment generating function for additive noise and stationary processes}
\noindent To evaluate the average and the variance of pointwise mutual information, we introduce the moment generating function:
\begin{equation}
\label{eqn:methods:MGF}
    M_\lambda = \langle e^{\lambda i_\tau} \rangle = \int d\bm{x}_\tau d\bm{x}_0 \, p(\bm{x}_{\tau}| \bm{x}_0) p(\bm{x}_0) \left(\frac{p(\bm{x}_{\tau}| \bm{x}_0)}{p(\bm{x}_\tau)} \right)^\lambda \;,
\end{equation}
such that $I_\tau = \partial_\lambda M_\lambda|_{\lambda=0}$ and $\mathcal{F}_\tau = \partial^2_\lambda \log M_\lambda |_{\lambda = 0} =  \partial^2_\lambda M_\lambda|_{\lambda=0} - (\partial_\lambda M_\lambda|_{\lambda=0})^2$, noting that $M_\lambda |_{\lambda = 0} = 1$. 

To make progress, we perform a Taylor expansion on $p_\tau$ up to the second order in $\bm{z} \equiv \bm{x}_\tau - \bm{x}_0$:
\begin{align}
    p_\tau^{-\lambda} = & p_0^{-\lambda} - \lambda \, p_0^{-\lambda-1} \,\bm{z} \cdot \bm{\nabla}p_0 + \\
    & + \lambda (\lambda + 1) p_0^{-\lambda-2} (\bm{\nabla} p_0)^T (\bm{z}^T\bm{z}) \,\bm{\nabla} p_0 + \nonumber \\
    & - \lambda \,p_0^{-\lambda-1} \,(\bm{z}^T\bm{z}) \cdot \bm{\nabla}^H p_0 \, \nonumber
\end{align}
where $\bm{\nabla}^H$ is the Hessian of $p_0$, i.e., $\bm{\nabla}^H_{ij} \,a = \partial_i \partial_j \,a$. By plugging this expression and Eq.~\eqref{eqn:methods:short_prop} into Eq.~\eqref{eqn:methods:MGF}, and restricting to the case of additive noise, we obtain:
\begin{widetext}
\begin{equation}
    M_\lambda = f^{(\lambda)} \int d\bm{x}_0 \, p_0^{-\lambda} \left( p_0 - \lambda \,\tau \,\bm{F}_x \cdot \bm{\nabla}p_0 - \frac{\lambda \, \tau}{1+\lambda} \left(\frac{1+\lambda}{p_0} (\bm{\nabla} p_0)^T \bm{D} \,\bm{\nabla} p_0 - \bm{D} \cdot \bm{\nabla}^H p_0 \right) \right)
\end{equation}
\end{widetext}
where $f^{(\lambda)}$ is reported in the main text. At stationarity, the moment generating function can be further simplified by considering that $p(\bm{x}_t) \to p_{\rm st}(\bm{x})$ and the probability flux $\bm{J}_{\rm st} = -\bm{F}_x p_{\rm st} + \bm{D} \cdot \bm{\nabla} p_{\rm st}$ is solenoidal, i.e., $\bm{\nabla} \cdot \bm{J}_{\rm st} = 0$. This allows us to write $\bm{D}\cdot \bm{\nabla}^H p_0 = p_0 \bm{\nabla} \cdot \bm{F}_x + \bm{F}_x \cdot \bm{\nabla} p_0$ and obtain the expression of the MGF in the main text. See the Supplementary Information for a step-by-step derivation.

\subsection{Gaussian processes}
\noindent In Gaussian processes, for which $\bm{F}(\bm{x}) = \bm{A} \bm{x}$, the system is described by its covariance matrix $\bm{S}$ which obeys the Lyapnuov equation $\bm{A}\bm{S} + \bm{S}\bm{A}^T = - 2 \bm{D}$. By using the Lyapunov equation, the entropy production can be written as \cite{nicoletti2024tuning}:
\begin{equation}
    \Sigma = \Tr\left[\bm{D}^{-1}\bm{A}\bm{S}\bm{A}^T\right] - \Tr\left[\bm{A}\right] \; .
\end{equation}
Similarly, the traffic can be written as
\begin{equation}
    \mathcal{T} = \frac{1}{4}\mathrm{Tr}\left[\bm{D}^{-1}\bm{A}\bm{S}\bm{A}^T\right] + \frac{1}{2}\mathrm{Tr}\left[\bm{A}\right]
\end{equation}
since $\langle \bm{A}^T\bm{x}^T\bm{D}^{-1}\bm{A}\bm{x}\rangle_{p_{\rm st}} = \mathrm{Tr}[\bm{D}^{-1}\bm{A}\bm{S}\bm{A}^T]$. The varentropy reduces to $\mathcal{F}_H[p_{\rm st}] = N/2$ for any Gaussian distribution. Substituting into Eq.~\eqref{eqn:F_additive}, one obtains $\mathcal{F}_\tau = N + 2\tau\,\mathrm{Tr}(\bm{A})$.

\bibliography{biblio}

\end{document}